\documentclass[%
 reprint,
 superscriptaddress,
 amsmath,amssymb,
 aps,
]{revtex4-2}

\usepackage{graphicx}% Include figure files
\usepackage{dcolumn}% Align table columns on decimal point
\usepackage{bm}% bold math
\usepackage{lipsum}
\usepackage{siunitx}
\usepackage{hyperref}
\usepackage{placeins}
\usepackage{etoolbox}
\usepackage{xcolor}
\usepackage{upgreek}
\usepackage{lineno}

\DeclareSIUnit\bar{bar}

\makeatletter
\def\@email#1#2{%
 \endgroup
 \patchcmd{\titleblock@produce}
  {\frontmatter@RRAPformat}
  {\frontmatter@RRAPformat{\produce@RRAP{*#1\href{mailto:#2}{#2}}}\frontmatter@RRAPformat}
  {}{}
}%

\begin{document}

\title{Resolving unconventional gap structure in kagome superconductors with hybrid microwave circuits }

\author{${\text{Yejin Lee}}^{\ddag}$ }
 \thanks{Corresponding author. email:yejin.lee@cpfs.mpg.de}
 \thanks{$^{\ddag}$These authors contributed equally to this work.}
 \affiliation{Max Planck Institute for Chemical Physics of Solids Dresden, 01187, Dresden, Germany}
 
\author {${\text{Haolin Jin}}^{\ddag}$}
 \affiliation{Max Planck Institute for Chemical Physics of Solids Dresden, 01187, Dresden, Germany}
 \affiliation{Institute of Solid State and Material Physics, Technische Universit{\"a}t Dresden, 01062 Dresden, Germany}

\author{Sushmita Chandra}
 \affiliation{Max Planck Institute for Chemical Physics of Solids Dresden, 01187, Dresden, Germany}

\author{Berit H. Goodge}
 \affiliation{Max Planck Institute for Chemical Physics of Solids Dresden, 01187, Dresden, Germany}

\author{Edouard Lesne}
 \affiliation{Max Planck Institute for Chemical Physics of Solids Dresden, 01187, Dresden, Germany}

\author{Tommaso Confalone}
\affiliation{Institute of Applied Physics, Technische Universit{\"a}t Dresden, 01062 Dresden, Germany}
\affiliation{Leibniz Institute for Solid State and Materials Research Dresden, 01069 Dresden, Germany}

\author{Francesco Tafuri}
 \affiliation{Department of Physics, University of Naples Federico II, Naples 80126, Italy}

\author{Davide Massarotti}
 \affiliation{Department of Electrical Engineering and Information Technology, University of Naples Federico II, Naples I-80126, Italy}

\author{Golam Haider}
 \affiliation{Leibniz Institute for Solid State and Materials Research Dresden, 01069 Dresden, Germany}
 
\author{Kornelius Nielsch}
 \affiliation{Institute of Applied Physics, Technische Universit{\"a}t Dresden, 01062 Dresden, Germany}
 \affiliation{Leibniz Institute for Solid State and Materials Research Dresden, 01069 Dresden, Germany}
 \affiliation{Institute of Materials Science, Technische Universit{\"a}t Dresden, 01062 Dresden, Germany}

\author{Bernd Büchner}
 \affiliation{Leibniz Institute for Solid State and Materials Research Dresden, 01069 Dresden, Germany}

\author{Claudia Felser}
 \affiliation{Max Planck Institute for Chemical Physics of Solids Dresden, 01187, Dresden, Germany}%Lines break automatically or can be 
 
\author{Debanjan Chowdhury}
 \affiliation{Department of Physics, Cornell University, Ithaca NY 14853, United States}

\author{Nicola Poccia}
 \affiliation{Leibniz Institute for Solid State and Materials Research Dresden, 01069 Dresden, Germany}
 \affiliation{Department of Physics, University of Naples Federico II, Naples 80126, Italy}

\author{Uri Vool}
  \thanks{uri.vool@cpfs.mpg.de}
 \affiliation{Max Planck Institute for Chemical Physics of Solids Dresden, 01187, Dresden, Germany}%Lines break automatically or can be forced with \\
 \affiliation{Leibniz Institute for Solid State and Materials Research Dresden, 01069 Dresden, Germany}

\date{\today}

\begin{abstract}

Unconventional superconductivity is a hallmark of exotic quantum matter, where determining the pairing symmetry is essential for uncovering its microscopic origin. 
Kagome superconductors provide a fertile landscape for emergent phenomena arising from strong electronic correlations and nontrivial band topology, yet their superconducting pairing symmetry remains elusive.  
The superconducting gap structure is commonly probed via electrodynamic response, but such measurements are inapplicable to thin flakes due to their small volume and delicate nature.
Superconducting microwave resonators offer a coherent and highly sensitive platform for probing electrodynamic responses, with versatile designs that enable incorporation of diverse materials and geometries. Here, we integrate flakes into microwave circuits, enabling noninvasive access to the superfluid response through contactless coupling that preserves structural integrity. By engineering the device geometry to suppress parasitic two-level-system losses that dominate dissipation in microwave circuits, we isolate the intrinsic material response. Remarkably, the temperature-dependent superfluid density exhibits linear behavior at low temperatures, consistent with a nodal gap structure. 
Our approach establishes a noninvasive platform for probing electrodynamic response of fragile superconducting flakes while advancing hybrid microwave architectures for quantum technologies.

\end{abstract}

 \maketitle

%\linenumbers

The discovery of new classes of unconventional superconductors has significantly expanded the landscape of quantum materials, spanning a wide range of correlated systems \cite{tsuei2000pairing, stewart2017unconventional,wang2024experimental}, as well as engineered structures \cite{amundsen2024colloquium, di2026van}. 
Among these, van der Waals (vdW) materials, including intrinsic layered compounds \cite{manzeli20172d, zhou2021superconductivity}, heterostructures \cite{geim2013van}, and twisted interfaces \cite{balents2020superconductivity,zhao2023time, yang2024twist} stand out as a unique low-dimensional setting for realizing correlated and topological phases of matter. 
Elucidating the superconducting gap structure is central to understanding the pairing mechanism. 
For exfoliated flakes, however, reduced dimensionality, limited sample volume, and delicate nature restrict the applicability of conventional bulk probes, motivating the development of sensitive, noninvasive approaches \cite{matsuda2006nodal, das2012reconstructing, yu2008evidence}.

Technological advances in circuit quantum electrodynamics \cite{blais2021circuit} over the past decade have opened a new frontier for sensing quantum materials. Superconducting microwave resonators are highly coherent, tunable devices operating at millikelvin temperatures and gigahertz frequencies ($\upmu$eV energy scale), providing a high-precision platform for probing the electrodynamic response of superconductors. This enables quantitative access to the kinetic inductance and superfluid density, which encode the low-energy excitation spectrum and constrain the pairing symmetry.
Coplanar waveguide architectures can be integrated with layered materials, and such hybrid circuits have been widely developed to investigate kinetic inductance \cite{kreidel2024measuring, kreidel2025observing, zaman2025kinetic, chistolini2025contactless}, and to realize new quantum devices \cite{schmidt2018ballistic, wang2019coherent, antony2021miniaturizing, sarkar2022quantum,wang2022hexagonal, butseraen2022gate, zollitsch2023probing, jin2025exploring, balgley2025coherent, blumenthal2026flux}. This approach is particularly well suited for accessing the electrodynamic properties of low-dimensional superconductors.

%%%
\begin{figure*}[t!]
   \center
   \includegraphics[width=0.85\textwidth]{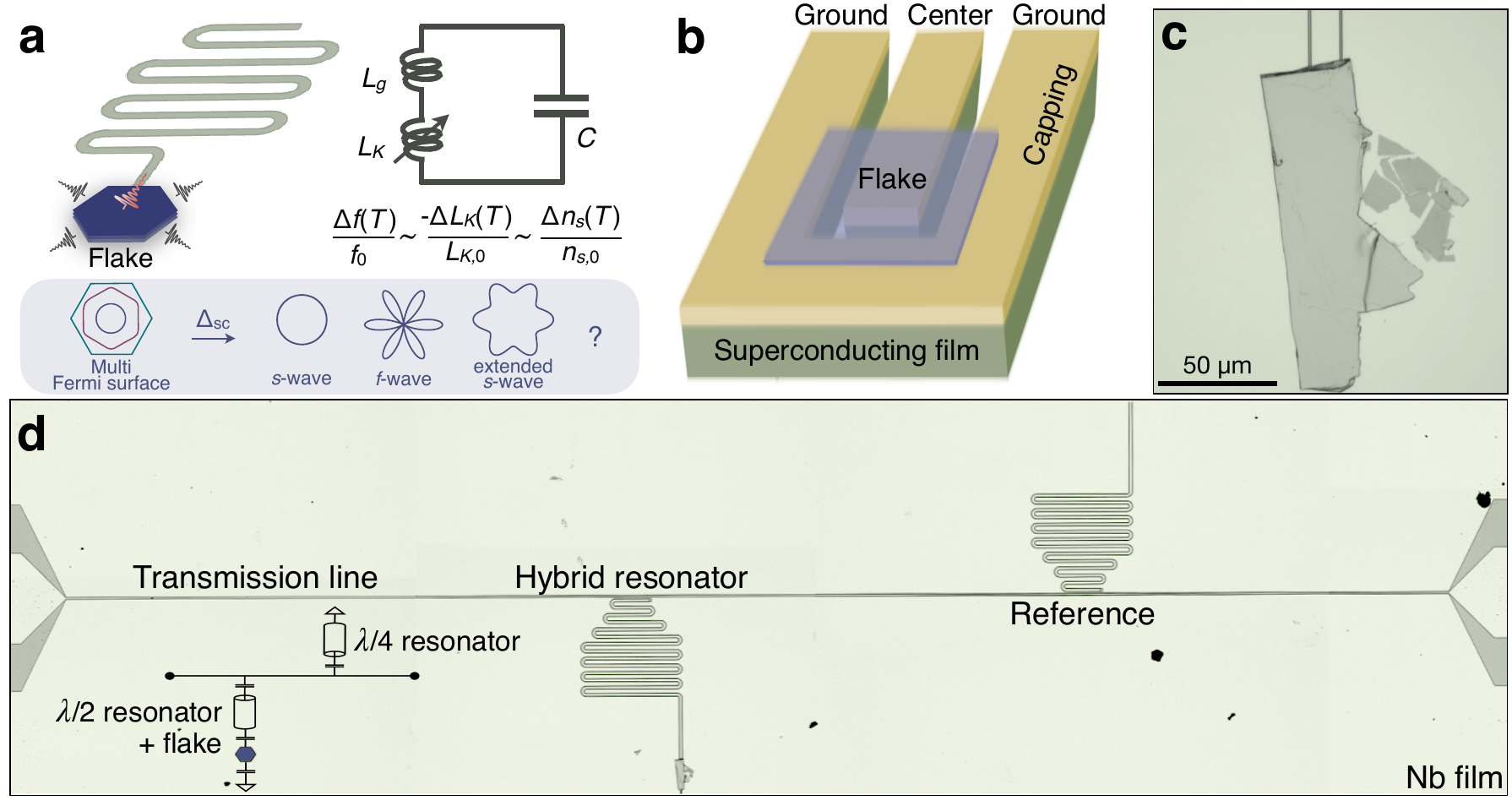}
   \caption{\textbf{Hybrid microwave resonator device configuration.} (a) Schematics for a hybrid microwave resonator coupled to a superconducting flake forming an effective LC oscillator. Kinetic inductance change manifests shifts in the resonance frequency, reflecting superfluid density response. Representative superconducting gap structures proposed for a kagome superconductor, including fully gapped, nodal and extended $s$-wave states, are illustrated. (b) Three-dimensional layout for the hybrid microwave device where a superconducting film is patterned into a coplanar design. A van der Waals (vdW) flake is placed between the center line and ground plane, on a surface capping layer grown on the superconducting film. (c) Optical micrograph of a representative vdW kagome flake integrated into the niobium resonator. (d) Overview of the hybrid microwave device comprising the hybrid resonator with the flake and a reference resonator capacitively coupled to the transmission line. The inset shows a circuit diagram corresponding to the device.
   }
   \label{fig:FIG1}
\end{figure*}
Realizing hybrid architectures based on vdW materials, however, remains challenging, especially for fragile layered compounds. 
Preserving their intrinsic properties while maintaining high-quality circuits is often nontrivial.
In such hybrid systems, surface and interfacial disorder arising from chemical processing, thermal exposure, or oxide formation can introduce parasitic dissipation via two-level systems (TLS) that couple to the electromagnetic field. While TLS provide a useful framework for understanding dissipation in superconducting circuits \cite{martinis2005decoherence, spiecker2023two}, their presence in hybrid devices, typically associated with amorphous interfacial dielectrics and surface imperfections \cite{woods2019determining, mcrae2020materials, mcrae2020dielectric}, can lead to decoherence and obscure the intrinsic response of embedded materials.
Integrating microwave circuits with thin flakes therefore necessitates preserving pristine crystals while minimizing TLS-induced dissipation. 

Our focus in this work is on kagome superconductors, where a key outstanding question concerns the excitation spectrum of their low-energy Bogoliubov quasiparticles. These materials are prominent quantum systems exhibiting intertwined electronic orders, including superconductivity, charge and pair density wave order, and time reversal symmetry-breaking \cite{wang2023quantum, wilson2024v3sb5, guo2025many,hu2024evidence,song2023anomalous,yu2021unusual, hao2022dirac}. These correlated phases are associated with the vanadium kagome lattice, which hosts multiple Dirac cones, flat bands, and van Hove singularities in its electronic band structure \cite{hao2022dirac, neupert2022charge}. 
The layered structure enables exfoliation into thin flakes, allowing exploration of the low-dimensional regime where superconducting properties exhibit pronounced thickness dependence \cite{song2021competition, wu2022nonreciprocal, song2023anomalous, zhang2024large}, making them an ideal platform for investigating pairing symmetry.
Experimental efforts on $\text{CsV}_{3} \text{Sb}_{5}$ have revealed diverse gap behaviors, including fully gapped \cite{zhong2023nodeless, mu2021s,duan2021nodeless, shan2022muon, gupta2022microscopic, sun2025clean}, anisotropic \cite{roppongi2023bulk, grant2024superconducting, kaczmarek2025direct} and nodal gaps \cite{hossain2025unconventional, mine2025observation}, as well as multiband \cite{xu2021multiband} and spectroscopic signatures of unconventional pairing \cite{zhao2021cascade, chen2021roton, deng2024chiral}. 
Collectively, these findings underscore the complexity of the superconducting state in $\text{CsV}_{3} \text{Sb}_{5}$.

Here, we employ a noninvasive integration method based on cryogenic dry transfer and capacitive, contactless coupling which preserves crystal integrity while enabling high-sensitive microwave measurements. We suppress the extrinsic TLS contribution in the hybrid device by optimizing the coupling geometry, thereby enhancing the inductive participation of the flake. This approach resolves the temperature dependence of the superfluid density down to the millikelvin regime, revealing a linear low-temperature evolution, indicative of a nodal gap structure. This establish a pathway to access the intrinsic superconducting electrodynamics of fragile vdW materials and provide a platform toward low-loss hybrid microwave devices based on vdW quantum materials. \\

\begin{figure*}[t!]
   \center
   \includegraphics[width=0.9\textwidth]{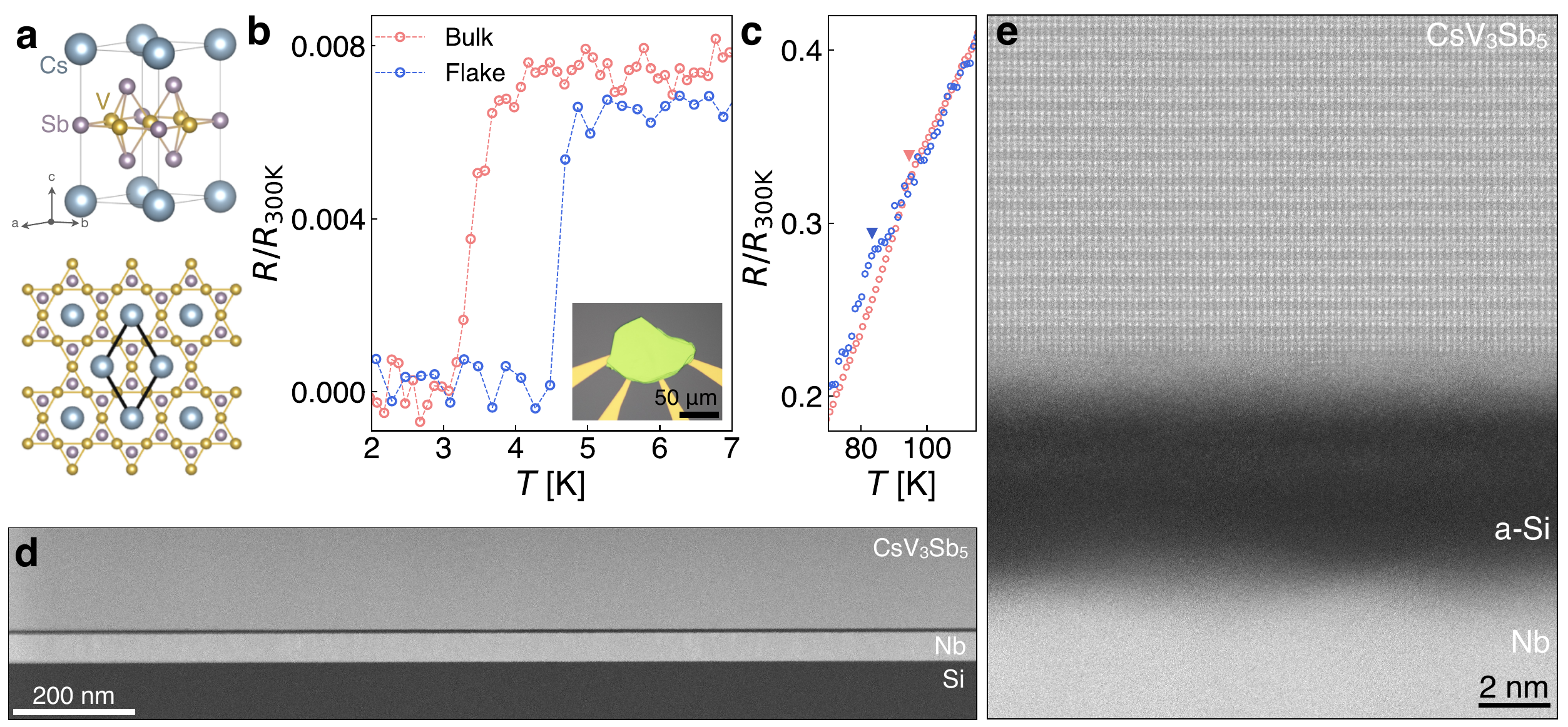}
   \caption{\textbf{Superconductivity in $\text{CsV}_3 \text{Sb}_5$ and device interface. }
   (a) Crystal structure of $\text{CsV}_3 \text{Sb}_5$ on the three dimensional plane and $ab$ plane, respectively. (b) Resistance as a function of temperature of the $\text{CsV}_3 \text{Sb}_5$ bulk crystal and the exfoliated 480 nm-thick flake, marked in red and blue, respectively. Inset: Optical micrograph of the flake transferred on the gold electrodes. 
   (c) Resistance vs. temperature measurements for the same devices in (b) at higher temperatures, indicating a charge density wave anomaly around 90 K. (d) Scanning transmission electron microscopic (STEM) image for the cross section of a prepared flake on the niobium film. (e) Closeup of the interface between the film and the flake showing the atomically resolved crystal structure of the $\text{CsV}_3 \text{Sb}_5$ flake.
   }
   \label{fig:FIG2}
\end{figure*}

\textbf{Hybrid microwave circuits}

We implement a hybrid microwave platform by integrating a superconducting flake into a superconducting coplanar-waveguide (CPW) resonator. The flake is placed at the current antinode of the resonator to maximize its electrodynamic participation (Fig. \ref{fig:FIG1}a). In this geometry, the flake contributes a kinetic inductance ($L_{k}$) to the total circuit inductance, with $L_k$ governed by the superfluid density ($n_s$). Consequently, temperature-dependent variations in $n_s$ manifest as shifts in the resonance frequency $f(T)$, enabling sensitive probing of superconducting electrodynamics and its low-energy excitations.

We focus on the kagome superconductor Cs$\text{V}_{3} \text{Sb}_{5}$, whose electronic structure comprises multiple Fermi-surface (FS) sheets \cite{mine2025observation}. The presence of several FS sheets suggests a band-dependent gap structure, illustrated schematically in Fig. \ref{fig:FIG1}a. The hybrid circuits are fabricated by transferring exfoliated Cs$\text{V}_{3} \text{Sb}_{5}$ flakes onto niobium CPW resonators using a cryogenic dry-transfer technique performed in an argon-filled glovebox (Methods). A schematic of the three-dimensional device architecture is shown in Fig. \ref{fig:FIG1}b, where the flake is capacitively coupled to the resonator through a thin silicon capping layer, forming a contactless interface. The device architecture consists of an open-ended $\lambda/2$ resonator terminated by the flake (Fig. \ref{fig:FIG1}c), alongside a reference $\lambda/4$ resonator with a directly grounded center conductor (Fig. \ref{fig:FIG1}d). Both resonators are capacitively coupled to a transmission line for microwave readout.

\begin{figure*}[t!]
   \center
   \includegraphics[width=0.87\textwidth]{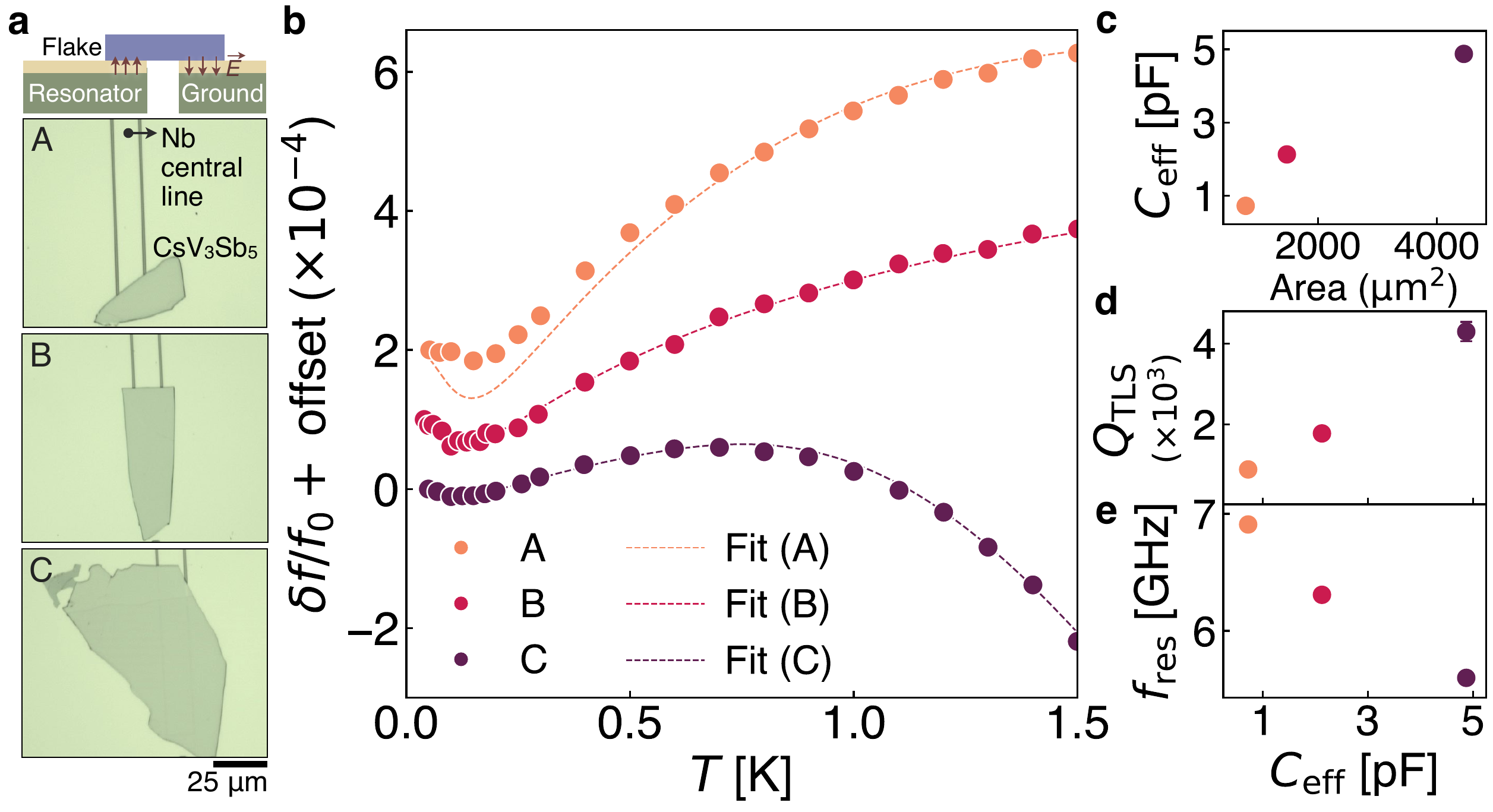}
   \caption{\textbf{Resonance frequency shift for devices of varying flake geometry.} (a) Optical micrograph of different hybrid devices (A-C) where a Cs$\text{V}_{3} \text{Sb}_{5}$ flake is integrated into the niobium resonator film. The top schematic shows a cross-sectional view of the device, displaying the flake is in partial contact with the central line and the ground across the isolation gap with the displacement field. The contact area of the flake on the resonator is increased from device A to C. (b) Temperature dependent resonance frequency shift for all devices, with the offset for visualization. The dashed line shows a model with the contribution from both a standard quasiparticle and two-level system bath ($\delta f/f_{0}$=$(\delta f/f_{0})_{\text{QP}}+(\delta f/f_{0})_{\text{TLS}}$). (c) Estimated effective capacitance $C_{\text{eff}}$, determined by the flake size and the dielectric capping layer, as a function of the contact area for all devices (A-C). (d) The extracted quality factor contributed from TLS ($Q_{\text{TLS}}$), and (e) measured resonance frequency at 50 mK as a function of $C_{\text{eff}}$. }
   \label{fig:FIG3new}
\end{figure*}
\textbf{Superconducting transitions}

To establish the superconducting properties and structural quality of the hybrid devices, we first examine the superconducting transitions in both a bulk crystal and a representative exfoliated flake.
The superconducting transition temperature ($T_{c}$) of $\text{CsV}_3 \text{Sb}_5$ has been observed to vary between 2.5 and 3.5 K \cite{ortiz2020cs, qian2021revealing, song2021competition, wu2022nonreciprocal}, reflecting sensitivity to surface termination and dimensionality. 
$\text{CsV}_3 \text{Sb}_5$ is a layered compound consisting of Cs spacer layers and V$-$Sb sheets formed by a vanadium kagome net (Fig. \ref{fig:FIG2}a).
Figure \ref{fig:FIG2}b displays the temperature dependence of the normalized resistance of the bulk and representative flake. Despite the flake thickness of 480 nm, approaching the bulk regime, the exfoliated flake exhibits a sharp superconducting transition at 4.5 K, higher than that of the bulk crystal (3 K) \cite{wu2022nonreciprocal,song2023anomalous,song2021competition}. 
In addition, the charge density wave (CDW) feature is observed at 80 K, reduced compared with its bulk counterpart (95 K) (Fig. \ref{fig:FIG2}c). 
The interplay of superconductivity and the CDW has been reported in a pressurized \cite{zhu2022double, zheng2022emergent} and uniaxial-strain applied crystals. In particular, an enhancement of the $T_c$ and reduction of $T_{\text{CDW}}$ have been observed under tensile strain \cite{qian2021revealing,lin2024uniaxial}, and may reflect modifications of interlayer coupling along the c-axis.  

To access the structural integrity in hybrid architectures, cross-sectional scanning transmission electron microscopy (STEM) is carried out on a similarly prepared flake.
Figure \ref{fig:FIG2}d reveals atomically flat layers without observable structural defects in the flake.
The sharp interface between the flake and the amorphous silicon-capped niobium film is preserved, without significant degradation at the bottom of the flake (see Fig. \ref{fig:FIG2}e), confirming well-defined interface essential for reliable microwave coupling.\\

\textbf{Microwave response in hybrid circuits}

We measure the temperature-dependent response of hybrid microwave circuits integrated with Cs$\text{V}_{3} \text{Sb}_{5}$ flakes with thicknesses in the several-hundred-nanometer regime. 
Surprisingly, the resonance frequency increases with increasing temperature. This anomalous upshift is commonly observed in superconducting circuits coupled to two-level-system (TLS) defects \cite{gao2008experimental, alexander2025power, jin2025exploring}, and can obscure the intrinsic response of the flakes.
To disentangle these effects from the intrinsic response, we design a series of devices with systematically varied flake geometries and increasing overlap area with the resonator, as shown in Fig.\ref{fig:FIG3new}a. 
We find that the magnitude of the frequency upshift decreases with increasing overlap area, indicating that TLS coupling is sensitive to the local electric-field distribution (Fig. \ref{fig:FIG3new}b).
To quantify the TLS contribution, we model the observed behavior as a sum of quasiparticle (QP) and TLS contributions \cite{gao2008physics}, with the total fractional frequency shift given by $\delta f/f_{0}$=$(\delta f/f_{0})_{\text{QP}}+(\delta f/f_{0})_{\text{TLS}}$ (see Supplementary Materials). 
The model reproduces the measured behavior well, and the extracted TLS-limited quality factor ($Q_{\text{TLS}}$) increases with flake area, indicating reduced TLS effects for larger interfaces. 
Considering the flake as a terminating capacitive load, we calculate the effective capacitance from its overlap with both the center conductor and ground plane (see Fig. \ref{fig:FIG3new}c). 
The resulting capacitance $C_{\text{eff}}$ increases approximately linearly with flake size, accompanied by a corresponding increase in $Q_{\text{TLS}}$, as plotted in Fig. \ref{fig:FIG3new}d. For smaller overlap areas, the reduced capacitance leads to a stronger concentration of the electric field at the interface, enhancing the coupling between the resonator field and interfacial TLS defects and producing a larger frequency shift. Consequently, larger flakes predominantly increase the effective capacitance, resulting in a lower base-temperature resonance frequency consistent with capacitive loading (Fig. \ref{fig:FIG3new}e).

\begin{figure*}[t!] 
   \center
   \includegraphics[width=0.7\textwidth]{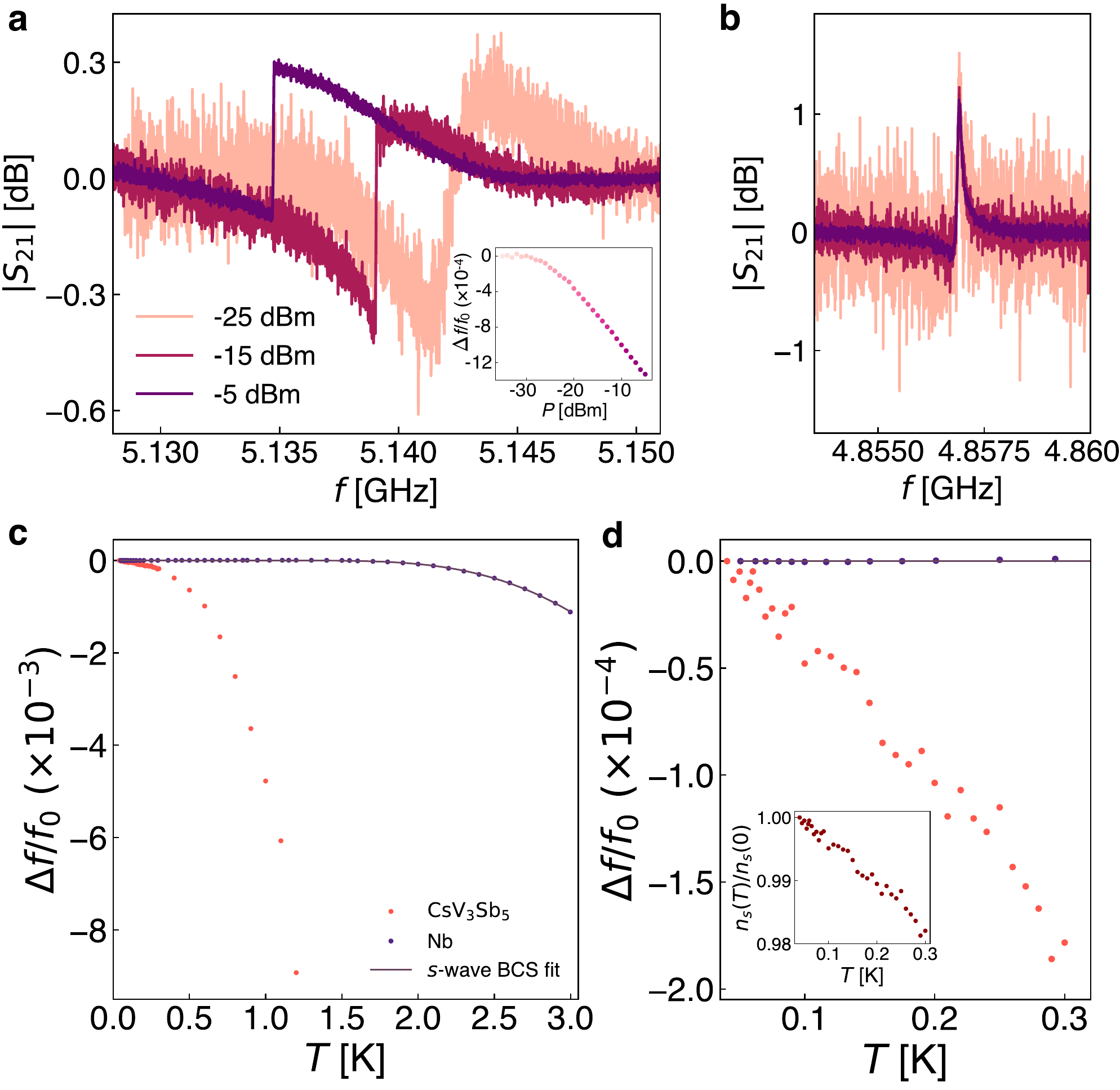}
   \caption{\textbf{Temperature dependent resonance frequency in hybrid device.} (a) Microwave transmission $S_{21}$ for the hybrid device, shown in Fig. 1c,d, measured at 50 mK under three different applied input powers and (b) the reference niobium resonator with $\lambda/4$ termination.  Inset of (a): the input power dependent frequency shift for the hybrid device. (c) The temperature dependence of the resonance frequency shift for the hybrid device and the niobium resonator with a model for a isotropic gap. (d) Zoom in for the temperature dependent resonance frequency shift at low temperatures. Inset of (d): Extracted relative superfluid density versus temperature of the corresponding hybrid device.}
   \label{fig:FIG4}
\end{figure*}

Throughout the investigation of different devices, increasing the overlap area enhances the effective capacitance and reduces electric-field concentration at interfacial regions, thus suppressing parasitic TLS contributions. This in turn improves sensitivity to the kinetic inductance response of the flakes. 
Well-defined planar interfaces and high crystal quality are therefore essential for isolating intrinsic behavior. Imperfect contact such as local gaps or inhomogeneous adhesion modifies the effective capacitance and current distribution, while introducing additional loss channels, leading to extrinsic features (See Supplementary Materials). 
Consistently, STEM measurements of an aged device exposed to ambient conditions for two months reveal degradation of the interface structure and indications of local crystal disorder, accompanied by increased dissipation and reduced quality factor.

In a following device, a larger overlap area yields an enhanced effective capacitance ($C_{\mathrm{eff}}=$8.68 pF). This device is shown in Fig. \ref{fig:FIG1}c,d.
Microwave transmission $S_{21}$ measurements of the hybrid device (see Fig. \ref{fig:FIG4}a) exhibit a distinct coherent mode with a quality factor of approximately $3\times10^3$, lower than that of the reference niobium resonator ($\approx 4.5\times10^4$) (Fig. \ref{fig:FIG4}b). The resonance frequency of the hybrid device exhibits a power-dependent shift toward low frequency with increasing input power, accompanied by a Duffing-like nonlinear response. In contrast, the reference resonator shows a power independent resonance frequency. Such power-dependent nonlinearities are often found in systems with high kinetic inductance \cite{valenti2019interplay, frasca2023nbn} and indicate the flake participates inductively in the hybrid circuit.
The resonance frequency of the hybrid device is comparable to that of the reference resonator, consistent with the flake acting as a capacitive termination while remaining inductively active through its kinetic inductance contribution. This capacitive coupling through the thin capping layer induces a slight deviation from an ideal $\lambda/4$ mode.

We further investigate the temperature dependence of the resonance frequency in the hybrid device.
The niobium reference resonator exhibits a temperature-independent response at low temperatures, followed by a gradual decrease near 3 K (Fig. \ref{fig:FIG4}c). This is consistent with Bardeen-Cooper-Schrieffer (BCS) theory for a conventional superconductor, with a fitted superconducting gap of 1.48 meV, in agreement with previous measurements of the niobium film \cite{pronin1998direct}. 
On the other hand, the hybrid device displays a markedly different evolution, with the resonance frequency decreasing approximately linearly from the lowest temperatures and a more pronounced drop above 0.5 K (Figure \ref{fig:FIG4}c,d). 
This behavior deviates from the expectation for an isotropic BCS gap and suggests the presence of low-energy quasiparticle excitations consistent with nodal superconductivity \cite{lee2006doping, hardy1993precision}. 
In such systems, the vanishing gap at nodal points on the Fermi surface leads to a linear temperature dependence of the superfluid density \cite{tsuei2000pairing,prozorov2006magnetic}. 
At higher temperatures, the response can be described by an effective energy scale of approximately 0.3 meV.
The pronounced low-temperature dependence is striking for a superconductor with $T_{c}$ of 4.5 K, and suggests a multiband scenario in which the smallest gap dominates the low-energy electrodynamics.
This is further supported by recent angle-resolved photoemission spectroscopy (ARPES) measurements reporting multiple superconducting gaps with distinct magnitudes, including a band with possible nodal or highly anisotropic characteristics \cite{mine2025observation}.

We extract the superfluid density $n_{s}(T)$ from the resonance frequency shift via the kinetic inductance $L_{\text{k}}$, obtained from the resonance condition using an effective capacitance, with $L_{\text{k}} \propto 1/n_{\text{s}}$ \cite{sutherland2003thermal} (see details in Supplementary Materials). 
The normalized superfluid density $n_s(T)/n_s(0)$ decreases with increasing temperature, exhibiting a weak low-temperature suppression consistent with anisotropic or nodal contributions. %and its weak reduction at low temperatures indicates nodal contributions, possibly in one band.
While this measurement is sensitive to low-energy quasiparticle excitations, it does not uniquely determine the superconducting order parameter, since the extracted $n_{s}(T)$ reflects a conductivity-weighted electrodynamic response rather than a momentum-resolved probe of the Fermi-surface. Accordingly, in multiband systems such as Cs$\text{V}_{3} \text{Sb}_{5}$, it imposes constraints on possible superconducting gap scenarios. 
To further quantify this regime, we analyze the slope of the linear regime of $n_s(T)/n_s(0)$ using expressions developed for cuprate superconductors \cite{lee1997unusual, sutherland2003thermal} (Supplementary Materials). The extracted slope lies near the lower bound of values obtained from previous experimental estimates \cite{ni2021anisotropic, grant2025superconducting, duan2021nodeless, gupta2022microscopic, kaczmarek2025direct, zhong2023testing, mine2025observation}, and is consistent with the overall experimental range. %range inferred in the literature. 
Operating in the $\upmu$eV energy scale, the present microwave technique probes the lowest-energy quasiparticle excitations, complementing ARPES measurements at meV energy scales \cite{mine2025observation}, and providing direct constraints on the low-energy electrodynamic response of the superconducting state. \\

%\section{Conclusion}

In summary, we have demonstrated a noninvasive platform for integrating the pristine van der Waals kagome superconductor $\text{CsV}_3 \text{Sb}_5$ into microwave resonators while preserving the structural integrity and maintaining a clean interface.
This approach enables high-coherence devices that sensitively detect changes in kinetic inductance in the vdW superconductor, advancing microwave circuits as probes of complex quantum materials.
We further show that engineering the geometry of the flake plays a crucial role in mitigating parasitic loss associated with TLS, thereby enabling access to the inherent electrodynamic response of the superconductor. The temperature dependence of the resonance frequency exhibits a linear decrease at low temperatures, consistent with a nodal superconducting gap structure. These results establish microwave resonators as a powerful platform for investigating correlated superconductivity in the low-dimensional limit and provide a route towards low-loss hybrid quantum technologies based on devices integrated with vdW quantum materials. \\

\noindent \textbf{Materials and Methods}

\subsection{Crystal synthesis}
High-quality single crystals of $\text{CsV}_3 \text{Sb}_5$ were synthesized using elemental cesium (liquid, Alfa Aesar, 99.98 \%), vanadium powder (Alfa Aesar, 99.9 \%), and antimony beads (Merck 99.999 \%). Due to the highly reactive nature of Cs, all preparation steps were carried out in an argon-filled glovebox with oxygen and moisture levels below 0.1 ppm. Crystal growth was performed using a self-flux method, in which a Cs–Sb binary eutectic mixture served as the flux \cite{ortiz2020cs, xiang2021twofold}. The starting materials were mixed in a molar ratio of $ \text{Cs}:\text{V}:\text{Sb} = 7:3:14$ and loaded in an alumina crucible, which was subsequently sealed in an evacuated quartz ampoule. The ampoule was heated to 1000 °C at a rate of \SI{50}{\celsius}/h and maintained at this temperature for 240 h. It was then slowly cooled to \SI{200}{\celsius} at a rate of \SI{3}{\celsius}/h, followed by cooling to room temperature. Excess flux was removed by rinsing the product with deionized water.

\subsection{Hybrid microwave resonator fabrication}
%(here, I: film fab, II: flake transfer)
Initially, a 60 nm niobium film is sputtered onto pure intrinsic silicon substrates. We then directly grow 8 nm silicon film as capping layer to prevent surface oxidation of niobium, causing dielectric loss, and proximity effect in galvanic contact. The microwave circuit design, including a transmission line and resonators, are patterned by electron-beam lithography, and followed with a reactive ion etching technique and oxygen plasma treatment. The resonator film is then loaded in argon gas filed glovebox. 
$\text{CsV}_3 \text{Sb}_5$ crystal is mechanically exfoliated on Si/Si$\text{O}_2$ substrates. The target flake is identified using optical microscope and the stage is cooled by liquid nitrogen. The flake is detached from the substrate using a pdms stamp at -80 celcius and then quickly transferred on the resonator film. More details of the device fabrication process can be found in ref. \cite{jin2025exploring}.

\subsection{Electronic transport measurements}
Electrical resistivity measurement on bulk single crystal was conducted using a Quantum Design Physical Property Measurement System (PPMS) in a conventional four-probe configuration. Electrical contacts were made using Pt wires and silver paste. The samples were oriented with the c-axis parallel to the applied magnetic field in the cryostat, and an excitation current of 5 mA was used for the measurement.

Exfoliated $\text{CsV}_3 \text{Sb}_5$ flakes are transferred on a Cr/Au electrodes sputtered on Si/Si$\text{O}_2$ substrates, using the same transfer method for hybrid microwave devices. The temperature dependent resistance for the flake is measured using a standard four wire method and a SR830 lock in amplifier with a physical property measurement system (Quantum design). 

\subsection{Scanning transmission electron microscopy measurements}
Cross-sectional lamellas of exfoliated $\text{CsV}_3 \text{Sb}_5$ were prepared using the standard focused ion beam (FIB) lift-out procedure on a Thermo Fisher Scientific Helios 5 CX FIB. Scanning transmission electron microscopy (STEM) images were collected in high-angle annular dark-field (HAADF) configuration on a double aberration-corrected JEOL ARM300F operating at 300 kV with a 30 mrad convergence angle. 

\subsection{Microwave measurements}
The microwave device is loaded in a Bluefors LD-400 dilution refrigerator with a base temperature of approximately 30 mK. Transmission spectra are acquired using a vector network analyzer (VNA, Keysight P9373A). The input microwave lines are equipped with attenuators at multiple cryogenic stages to suppress thermal noise. On the output side, a high-electron-mobility transistor (HEMT) amplifier (LNF) is thermally anchored to the 4 K stage, and a circulator at the Still stage is employed to isolate the device from amplifier backaction. At room temperature (300 K), the output signal is further amplified using a MITEQ amplifier. The sample temperature is controlled via a local heater and a thermometer near the device using Lakeshore 372 controller. In the temperature-dependent measurements, the system is allowed to stabilize for several minutes at each temperature for recording the transmission spectrum. \\

{}

%\newpage

\noindent \textbf{Acknowledgments}: We are grateful to Ioan Pop, Philip Moll, Cissy Suen, Haijing Zhang and Andrew Mackenzie for fruitful discussions. We thank to Ronny Engelhart, Martin Bauer, Sandra Nestler and Tino Wolf for technical support. \textbf{Funding}: The work is funded by the European Union (ERC-StG, cQEDscope, 101075962, and ERC-CoG, 3DCuT, 101124606) and partially supported by the Deutsche Forschungsgemeinschaft (DFG, German Research Foundation): DFG 539383397, DFG 460444718, DFG 512734967, DFG 452128813, DFG 539383397, DFG 572638824. F.T. and D. M. would like to acknowledge the PNRR MUR Project PE0000023 NQST. D.C. is supported in part by a grant from the Department of Energy (DE-SC0026112) under the Early Career Research Program. S.C. and C.F. acknowledge financial support by the DFG through QUAST (project ID FOR 5249) and Würzburg-Dresden Cluster of Excellence ctd.qmat - Complexity, Topology and Dynamics in Quantum Matter (EXC 2147, project ID 390858490). \noindent \textbf{Author contributions}: Y.L. performed the integration of flakes onto the resonators. H.J. fabricated the resonators. Y.L. and H.J. carried out the microwave measurements. H.J. developed the theoretical model. Y.L. H.J. and U.V. performed analysis with assistance of B.G. and D.C.. S.C. provided the crystals and performed bulk measurements. B.G. carried out transmission electron microscopy measurements. E.L. sputtered the superconducting films. Y.L. prepared the initial draft and wrote the manuscript with input from all authors. U.V. is a principal investigator. \noindent\textbf{Competing interest}: The authors declare that they have no competing interests. \noindent\textbf{Data and materials availability}: All data needed to evaluate the conclusions in the paper are present in the paper and/or the Supplementary Materials. \\

\end{document}

\newpage
\setcounter{secnumdepth}{3}

\onecolumngrid
%\appendix
\begin{center}
\large{\textbf{Supplementary Materials for \\ 
``Resolving unconventional gap structure in kagome superconductors \\ 
with hybrid microwave circuits''}}
\end{center}

\author{${\text{Yejin Lee}}^{*, \ddag}$ }
% \thanks{Corresponding author. email:yejin.lee@cpfs.mpg.de}
 \affiliation{Max Planck Institute for Chemical Physics of Solids Dresden, 01187, Dresden, Germany}

\author {${\text{Haolin Jin}}^{\ddag}$}
 \affiliation{Max Planck Institute for Chemical Physics of Solids Dresden, 01187, Dresden, Germany}
 \affiliation{Institute of Solid State and Material Physics, Technische Universit{\"a}t Dresden, 01062 Dresden, Germany}

\author{Sushmita Chandra}
 \affiliation{Max Planck Institute for Chemical Physics of Solids Dresden, 01187, Dresden, Germany}

\author{Berit H. Goodge}
 \affiliation{Max Planck Institute for Chemical Physics of Solids Dresden, 01187, Dresden, Germany}

\author{Edouard Lesne}
 \affiliation{Max Planck Institute for Chemical Physics of Solids Dresden, 01187, Dresden, Germany}

\author{Tommaso Confalone}
\affiliation{Institute of Applied Physics, Technische Universit{\"a}t Dresden, 01062 Dresden, Germany}
\affiliation{Leibniz Institute for Solid State and Materials Research Dresden, 01069 Dresden, Germany}

\author{Francesco Tafuri}
 \affiliation{Department of Physics, University of Naples Federico II, Naples 80126, Italy}

\author{Davide Massarotti}
 \affiliation{Department of Electrical Engineering and Information Technology, University of Naples Federico II, Naples I-80126, Italy}

\author{Golam Haider}
 \affiliation{Leibniz Institute for Solid State and Materials Research Dresden, 01069 Dresden, Germany}
 
\author{Kornelius Nielsch}
 \affiliation{Institute of Applied Physics, Technische Universit{\"a}t Dresden, 01062 Dresden, Germany}
 \affiliation{Leibniz Institute for Solid State and Materials Research Dresden, 01069 Dresden, Germany}
 \affiliation{Institute of Materials Science, Technische Universit{\"a}t Dresden, 01062 Dresden, Germany}

\author{Bernd Büchner}
 \affiliation{Leibniz Institute for Solid State and Materials Research Dresden, 01069 Dresden, Germany}

\author{Claudia Felser}
 \affiliation{Max Planck Institute for Chemical Physics of Solids Dresden, 01187, Dresden, Germany}%Lines break automatically or can be 
 
\author{Debanjan Chowdhury}
 \affiliation{Department of Physics, Cornell University, Ithaca NY 14853, United States}

\author{Nicola Poccia}
 \affiliation{Leibniz Institute for Solid State and Materials Research Dresden, 01069 Dresden, Germany}
 \affiliation{Department of Physics, University of Naples Federico II, Naples 80126, Italy}

\author{Uri Vool}
 \affiliation{Max Planck Institute for Chemical Physics of Solids Dresden, 01187, Dresden, Germany}
 \affiliation{Leibniz Institute for Solid State and Materials Research Dresden, 01069 Dresden, Germany}

\maketitle
\onecolumngrid

\makeatletter
\renewcommand{\fnum@figure}{\textbf{Fig.\thefigure}}
\makeatother
\renewcommand{\thefigure}{S\arabic{figure}}
%\newpage
\renewcommand{\thetable}{S\arabic{table}}\makeatletter
\renewcommand{\fnum@table}{\textbf{Table~\thetable}}
\makeatother

\tableofcontents

\newpage

\section{S\MakeLowercase{ingle crystal X-ray diffraction }}
\begin{figure*}[h!]
   \centering
   \includegraphics[width=0.5\textwidth]{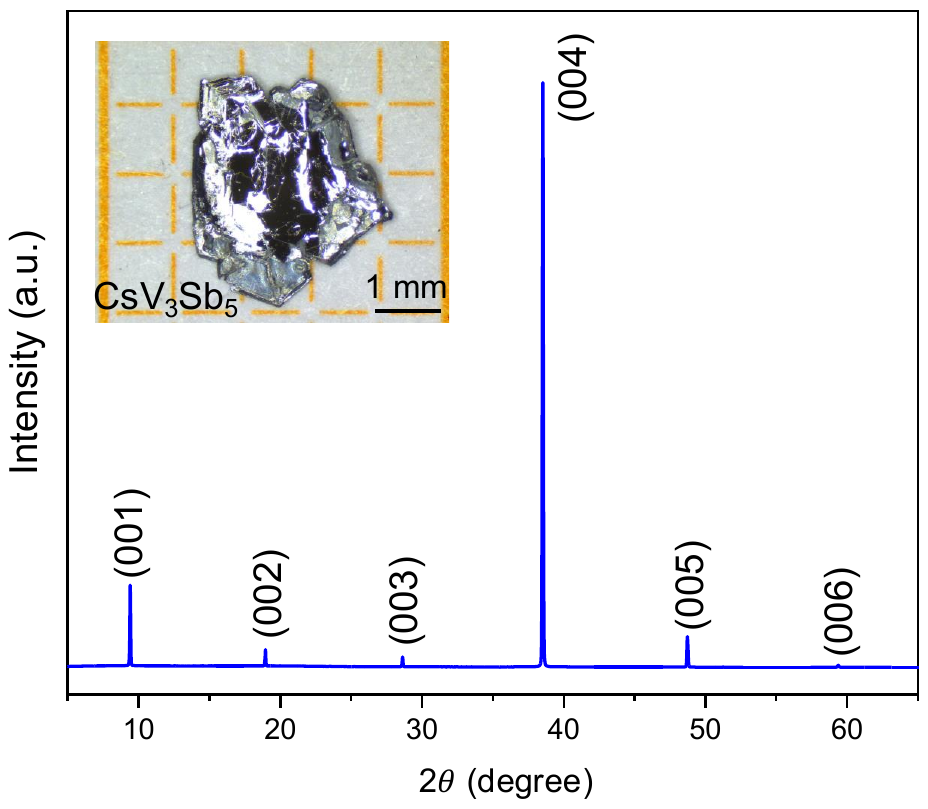}
   \caption{XRD pattern. The phase-purity of the $\mathrm{CsV}_{3}\mathrm{Sb}_{5}$ crystals was confirmed by X-ray diffraction (XRD). The presence of sharp (00l) reflections indicates high crystallinity and a preferred c-axis orientation. The plate-like crystals exhibit shiny surfaces with a well-defined hexagonal morphology. The typical dimensions of the crystals are $\sim3 \times3 \times 0.5$ $\text{mm}^3$. }
   \label{fig:supp_TEM}
\end{figure*}

\section{T\MakeLowercase{ransmission electron microscopy characterization for a hybrid device}}
\begin{figure*}[h!]
   \centering
   \includegraphics[width=0.89\textwidth]{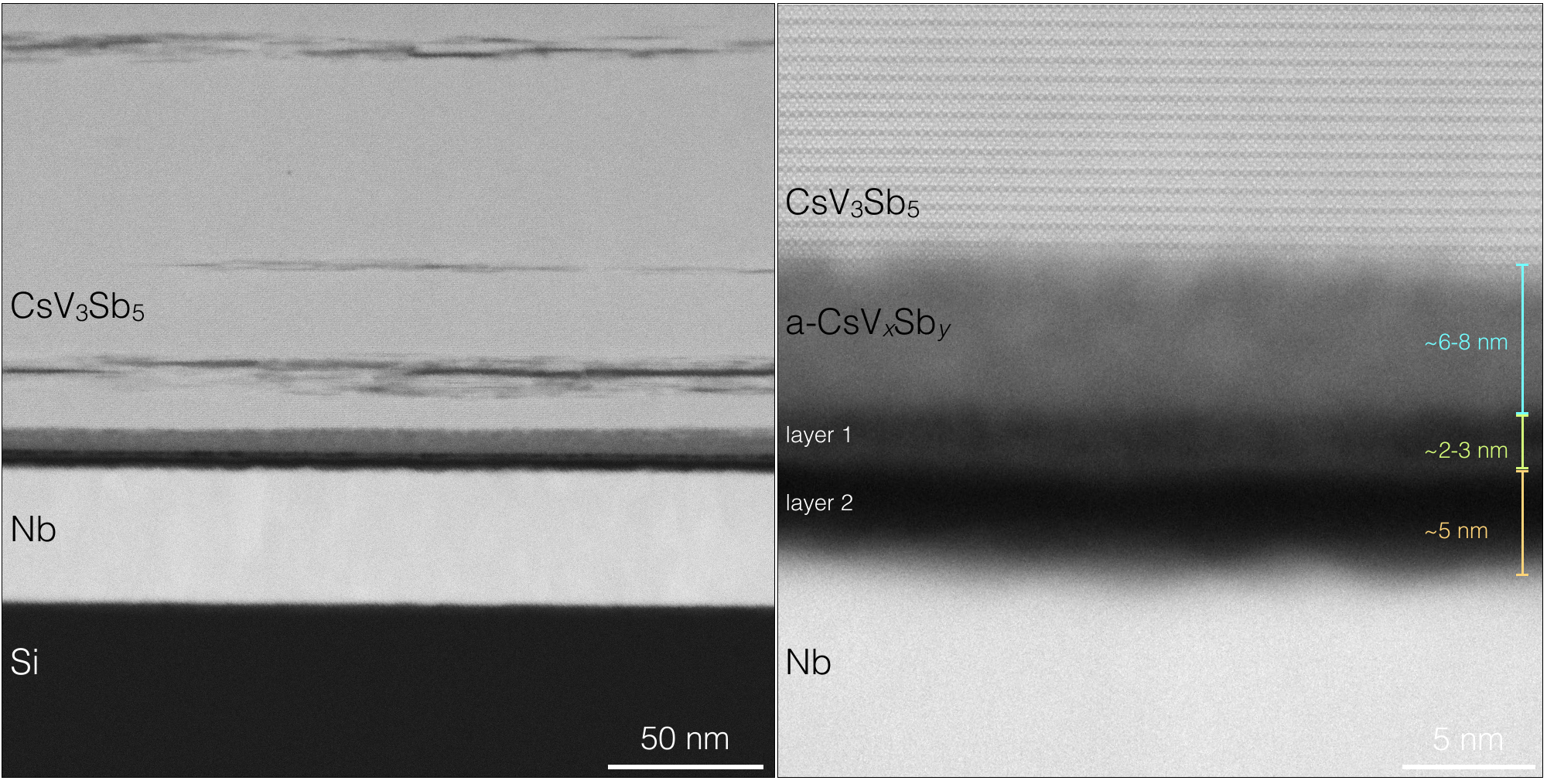}
   \caption{Scanning Transmission Electron Microscopy (STEM) image for the cross section in the aged hybrid device (left) and closeup at the interface (right).  }
   \label{fig:supp_TEM}
\end{figure*}

We performed scanning transmission electron microscopy (STEM) measurement for a hybrid device integrated with a $\mathrm{CsV}_{3}\mathrm{Sb}_{5}$ flake that had been exposed to ambient conditions. The flake was transferred onto the superconducting film using the same procedure as that used for the devices discussed in the main text. We observe delaminated stacking defects and non-uniform and destructive interface between the niobium/silicon film and the flake (\ref{fig:supp_TEM}). The bottom surface of the flake appears substantially more degraded than that of a freshly transferred flake (see Fig. 2 in the main text), exhibiting extensive amorphization. In the silicon capping layer, the top surface shows significant interfacial degradation where it contacts the bottom of the flake. The microwave response of this device fluctuates with temperatures and exhibits a markedly low quality factor.

\vspace{2em}
\section{E\MakeLowercase{ffective capacitance estimation}}

\subsection{Model of the transmission line resonator}
\begin{figure*}[h!]
   \centering
   \includegraphics[width=0.8\textwidth]{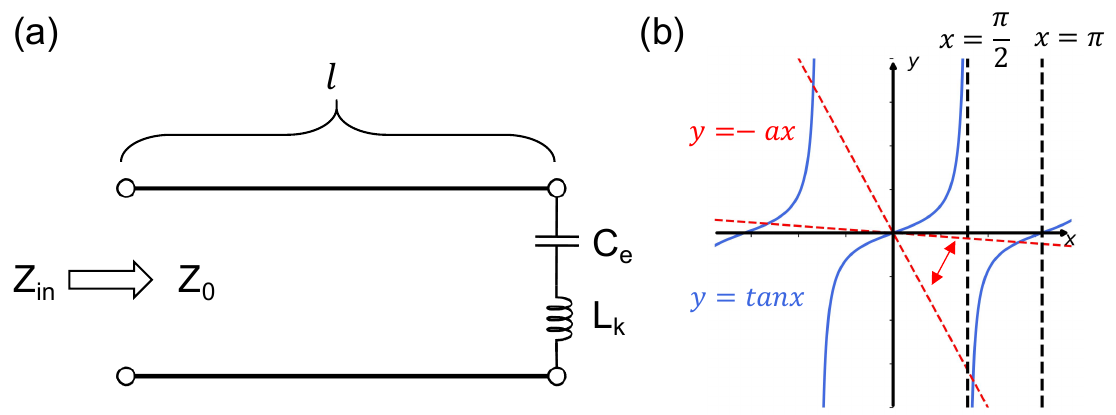}
   \caption{(a) Schematic of the terminated transmission-line resonator coupled to a flake as an inductive-capacitive element ($L_k$, $C_e$). (b) Numerical solution of $-a x = \tan x$ for different $a$ values, showing the resonance shift between the $\lambda/2$ and $\lambda/4$ modes.}
   \label{fig:FIGS2}
\end{figure*}

In our hybrid device configuration, the flake is positioned at a half-wavelength ($\lambda/2$) resonator and effectively terminates to ground. To model this system, we describe the $\lambda/2$ resonator as an open-ended transmission line of length $l$, terminated by a load impedance composed of the kinetic inductance $L_{k}$ and effective capacitance $C_{e}$ (Fig. \ref{fig:FIGS2}(a)).

The input impedance of a transmission-line resonator terminated by a load impedance is given by
\begin{equation}
Z_{\mathrm{in}}(\omega) = Z_{0} \frac{Z_{L} + j Z_{0} \tan (\beta l)}{Z_{0} + j Z_{L} \tan (\beta l)},
\end{equation}
with
\begin{equation}
    \beta=\omega / v_{p},
\end{equation}
where $v_{p} = 1/\sqrt{L'C'}$ is the phase velocity and $L'$, $C'$ are the inductance and capacitance per unit length. $Z_{0}$ and $Z_{L}$ denote the characteristic impedance of the transmission line and the terminating load, respectively. 
The propagation constant is determined using the calibrated $v_{\text{p}}$, extracted from the reference $\lambda/4$ resonator frequency.
Resonances occurs when $Z_{\mathrm{in}} \to \infty$, leading to the condition
\begin{equation}
Z_0 + j Z_L \tan(\beta l)=0.
\end{equation}
The load impedance of the flake is
\begin{equation}
Z_L(\omega)=j\omega L_k+\frac{1}{j\omega C_e}
= j\left(\omega L_k-\frac{1}{\omega C_e}\right).
\qquad
%\beta=\frac{\omega}{v_p}.
\end{equation}

Substituting Eq.(3) into Eq.(2) yields the general resonance condition
\begin{equation}
\tan(\beta l)=\frac{Z_0}{\omega L_k-\dfrac{1}{\omega C_e}}.
\end{equation}

For flakes with lateral dimensions much smaller than the effective wavelength ($\beta l_{\text{flake}} \ll 1$), the termination can be modeled as a lumped capacitance with impedance $Z_{L} = 1/(j \omega C_{e})$. In this limit, Eq. (4) reduces to
\begin{equation}    
-Z_{0}\omega C_{e} = \tan(\beta l).
\end{equation}
Defining the dimensionless variables $x=\beta l$ and $a = v_{p}Z_{0}C_{e}/l = C_{e}/(C'l)$, Eq.(8) becomes
\begin{equation} 
- a x = \tan x.
\end{equation}
This equation can be solved numerically, as shown in Fig.~\ref{fig:FIGS2}(b).

\vspace{2em}
\subsection{Estimation of effective capacitance $C_{\text{e}}$}

\begin{figure*}[h!]
   \centering
   \includegraphics[width=0.4\textwidth]{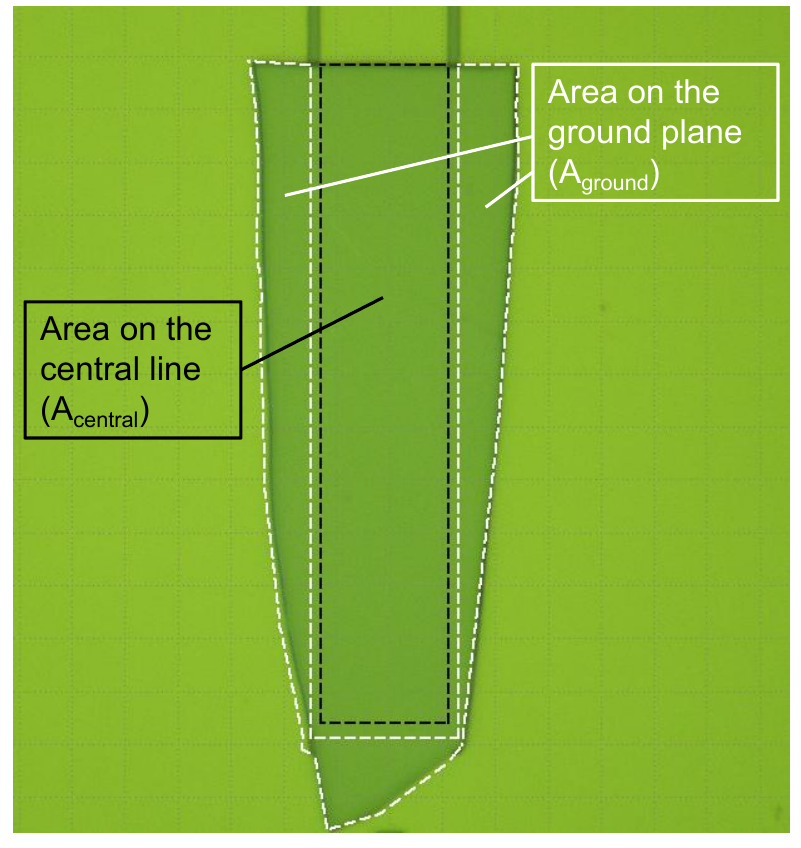}
   \caption{Schematic of the contact capacitor for the vdW flake on the central line and on the ground plane of the resonator.}
   \label{fig:Area_central}
\end{figure*}

The effective capacitance $C_{\text{e}}$ is determined from the device geometry using a parallel-plate approximation. This parameter is essential for capturing the contribution of the kinetic inductance $L_{\text{k}}$ to the resonance response.
Since the flake overlaps both the central conductor and the ground plane, the total capacitance is modeled as a series combination:
\begin{equation}
    \frac{1}{C_{\mathrm{e}}} = 
    \left(
        \frac{1}{C_{\mathrm{ground}}}
        + 
        \frac{1}{C_{\mathrm{central}}}
    \right),
\end{equation}
where the capacitance is associated with the central line ($C_\mathrm{cerntral}$) and the ground plane ($C_\mathrm{ground}$). Each contribution is estimated using
\begin{equation}
    C_{} = \frac{\varepsilon_{r}\varepsilon_{0}A_{\mathrm{}}}{d},
\end{equation}
where $A$ is the contact area between the flake and either the central conductor or the ground plane (Fig.~\ref{fig:Area_central}), $d$ is the separation between the resonator and the flake, $\varepsilon_{0}$ is the vacuum permittivity, and $\varepsilon_{r}$ is the effective relative permittivity. Base on STEM measurements (Fig.2), effective permittivity is determined by the dielectric stack at the interface. We model the interface by considering an intrinsic silicon layer thickness $t_{\mathrm{Si}}$ and a top silicon oxide layer of thickness $t_\mathrm{SiO_{\textit{x}}}$, which together define the effective dielectric environment.
\begin{equation}
\frac{d}{\varepsilon_{r}} = \frac{t_{\text{Si}}}{\varepsilon_{\text{Si}}}+\frac{t_{\text{ SiO}_x}}{\varepsilon_{\text{ SiO}_x}}
\end{equation}
The resulting values of $C_{\text{e}}$ for all devices are summarized in Table \ref{tab:full_kagome_table}. The effective permittivity is calibrated using the reference  \(\lambda/4\) resonator frequency to account for variations arising from the fabrication process.

\begin{table}[h]
\centering
\small
\begin{tabular}{lrrlrrrrr}
\toprule
Device &
area ($\upmu$m$^2$)&
area$_{\mathrm{central}}$ ($\upmu$m$^2$)&
 $C_{e}$(pF)&$f_{\lambda/4}$ (GHz) &
$v_p$ (m/s) &
thickness (nm) &
$f_{\mathrm{hyb}}$ (GHz) &
$Q_{\mathrm{int}}$ \\
\midrule
A & 780 & 138 &  0.7259&4.89 & 1.095$\times10^8$ & 200 & 6.9087 & 12510 \\
B & 1474 & 728 &  2.1295&4.9163 & 1.10$\times10^8$ & 328 & 6.3080 & 21507 \\
C & 4456 & 975 &  4.8735&4.7892 & 1.07$\times10^8$ & 111 & 5.5991 & 10680 \\
D & 5377 & 1580 &  8.6844&4.8569 & 1.09$\times10^8$ & 410 & 5.1412 & 2803 \\
\end{tabular}
\caption{Device parameters to estimate $C_e$.} The reference resonator design is identical for devices A-D. The devices A-C are the corresponding devices shown in Fig. 3 and the device D is discussed in Fig.2 and Fig.4 in the main text.
\label{tab:full_kagome_table}
\end{table}

\vspace{2em}
\subsection{Limiting behavior of the resonance condition}

To understand how $C_{e}$ affects the resonance spectrum, we analyze Eq. (9) in two limiting regimes:

\subsubsection{Small $C_{e}$ ($a \ll 1$, approaching $\lambda/2$ mode)}

For $x = \pi-\delta$ with $\delta \ll 1$, we use $\tan(x) \approx -\delta$. Substituting into Eq.(9) gives 
\begin{equation}
 a(\pi - \delta) = \delta,
\end{equation}
which leads $\delta \approx a \pi$. Therefore, 
\begin{equation}
f = f_{\lambda/2}\left(1 - \frac{C_{e}}{C'l}\right).
\end{equation}
In this regime, the system remains close ot the half-wavelength resonance with a small capacitive element.

\subsubsection{Large $C_{e}$ ($a \gg 1$, approaching $\lambda/4$ mode)}

For $x= \pi/2 +\delta$ with $\delta \ll1$, we use $\tan(x) \approx -1/\delta$.  Substituting into Eq.(9) gives 

\begin{equation}
a\left(\frac{\pi}{2} + \delta\right) =  \frac{1}{\delta} .
\end{equation}
With $a \gg 1$, this yields $\delta \approx 2 / (a\pi)$. The resonance frequency becomes
\begin{equation}   
f = f_{\lambda/4}\left(1 + \frac{4C'l}{C_{e}\pi^{2}}\right).
\end{equation}
In this regime, the resonance frequency evolves toward the quarter-wavelength mode as capacitive loading increases.

\vspace{2em}
\section{T\MakeLowercase{emperature dependent resonance frequency of hybrid devices with different flake geometry}}

\begin{figure*}[h!]
   \centering
   \includegraphics[width=0.99\textwidth]{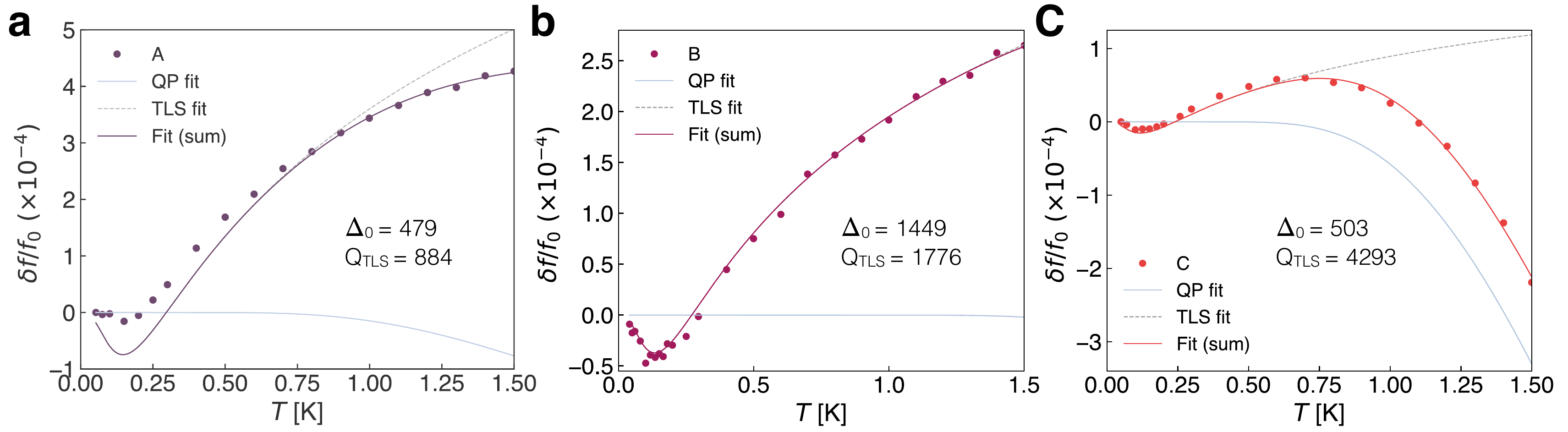}
   \caption{The temperature dependence of the resonance frequency for different hybrid devices A-C. Solid line displays the isotropic BCS-fit, and dashed line exhibits two level system effect on temperature dependent frequency.}
   \label{fig:multi_fit_TLS}
\end{figure*}

In this section, we provide a comprehensive description of the fitting method that were used to analyze the temperature dependent resonance frequency of hybrid devices A-C, presented in the main text. In these devices, we observe an anomalous upshift in frequency while increasing temperature, which is often found in two level system (TLS) coupled materials. We combine the standard quasiparticle (QP) model with TLS contribution, where the total fractional frequency shift is given by a combination of contributions from both the QP and TLS [68]:% \cite{gao2008physics}:

\begin{align}
\frac{\delta f(T)}{f_{0}} &= \left(\frac{\delta f(T)}{f_{0}}\right)_{\mathrm{QP}} + \left(\frac{\delta f(T)}{f_{0}}\right)_{\mathrm{TLS}}
\end{align}
\begin{align}
\left(\frac{\delta f(T)}{f_{0}}\right)_{\mathrm{QP}}=A \sqrt{\frac{2 \pi \Delta_{0}}{T}} \mathrm{e}^{-\Delta_{0} / T}
\end{align}
\begin{align}
\begin{split}
\left(\frac{\delta f(T)}{f_{0}}\right)_{\mathrm{TLS}} =
\frac{1}{\pi Q_{\mathrm{TLS}}} \operatorname{Re}\biggl[\Psi\left(\frac{1}{2}+i \frac{\hbar \omega}{2 \pi k_{B} T}\right) 
-\ln \left(\frac{\hbar \omega}{2 \pi k_{B} T}\right)\biggr]
\end{split}
\end{align}

For each device (A-C), the experimental data were fitted to the model, and the following parameters were extracted: gap size ($\Delta_0$) and the quality factor of TLS ($Q_{\text{TLS}}$). The fitted curves, along with the corresponding $\Delta_0$ and $Q_{\text{TLS}}$, are shown in Fig.\ref{fig:multi_fit_TLS}.

\section{I\MakeLowercase{mpact of terminating capacitance on two level system phenomena}}

The resonance frequency of a device is strongly influenced by the effective terminating capacitance $C_{\text{e}}$. In addition, TLS defects in the dielectric environment contribute to the effective relative permittivity, and consequently induce anomalous temperature-dependent shifts in the resonance frequency. This leads to a localized electric-field distribution that modifies the resonator response. The dielectric response associated with TLS defects can be expressed as [68]: %\cite{gao2008physics}:

\begin{equation}
\epsilon_{\mathrm{TLS}}(\omega)
= -\frac{2 P d_{0}^{2}}{3}
\left[
\Psi\!\left(
\frac{1}{2}
+ i\frac{\hbar \omega -i \Gamma}{2  \pi k_{B} T}
\right)
- \ln\!\left(
\frac{\varepsilon_{\max}}{2 \pi k_{B} T}
\right)
\right],
\end{equation}

where \(P\) is the density of two-level states, \(d_{0}\) is the electric dipole moment, \(\Psi\) is the digamma function, \(\Gamma\) is the TLS linewidth, and \(\varepsilon_{\max}\) represents the high-energy cutoff of the TLS distribution.

Based on this dielectric response, the total terminating capacitance is then given by:
\begin{equation}
C_{e} = C_{e0} + {C_{\mathrm{TLS}}} 
      = C_{e0}\bigl(1 + F\delta_{\mathrm{TLS}}(T)\bigr)
      = C_{e0}\!\left(1 + F\frac{\epsilon_{\mathrm{TLS}}(T)}{\epsilon_{\text{host}}}\right),
\end{equation}

where $C_{e0}$ is the nominal capacitance in the absence of TLS, $\delta_{\mathrm{TLS}}(T) = 2\pi P d_{0}^{2}/(3\epsilon_{\text{host}})$, $ \epsilon_{\text{host}}$ is the permittivity of the host dielectric, and $F$ is the filling factor. The $F$ accounts for the fact that the TLS-hosting material, occupying a volume $V_{h}$, may constitute only a fraction of the total resonator volume $V$, thereby reducing its contribution to the resonance frequency shift. The term $C_{\text{TLS}}$ quantifies the modification of the effective capacitance arising from the temperature-dependent dielectric response of TLS defects.
Assuming a small perturbation, such that $\frac{1}{1+x}\simeq 1-x$ for $|x| \ll1$, the fractional frequency shift induced by TLS effects can be written as

\begin{equation}
\frac{\Delta f(T)}{f}
\approx
-\frac{4 C' l}{\pi^{2} C_{e0}}\,F\delta_{\mathrm{TLS}}(T),
\end{equation}

where $C'$ is the geometry capacitance per-unit-length and $l$ is the resonator length. 
Combining these expressions yields

\begin{equation}
\frac{\Delta f(T)}{f}
\approx
\frac{4}{\pi^{2}}
\frac{C' l}{C_{e0}}
F\delta_{\mathrm{TLS}}(T)
\operatorname{Re}\biggl[\Psi\left(\frac{1}{2}+i \frac{\hbar \omega}{2 \pi k_{B} T}\right) 
-\ln \left(\frac{\hbar \omega}{2 \pi k_{B} T}\right)\biggr]
\end{equation}

The equation~(21) describes the temperature-dependent dispersive frequency shift arising from TLS defects coupled through the terminating capacitor. The magnitude of the frequency shift scales with the geometric capacitance $C'$ of the niobium resonator, and is inversely proportional to the terminating capacitor $C_{e0}$.

Consequently, the terminating capacitance provides a convenient tuning parameter for controlling the coupling to TLS defects and, therefore, the resulting temperature-dependent frequency shift. This design strategy enables optimizatiion of the device performance by reducing the impact of TLS-induced dielectric fluctuations.

\vspace{2em}
\section{T\MakeLowercase{emperature dependent quality factor }}  

\begin{figure*}[h!]
   \centering
   \includegraphics[width=0.5\textwidth]{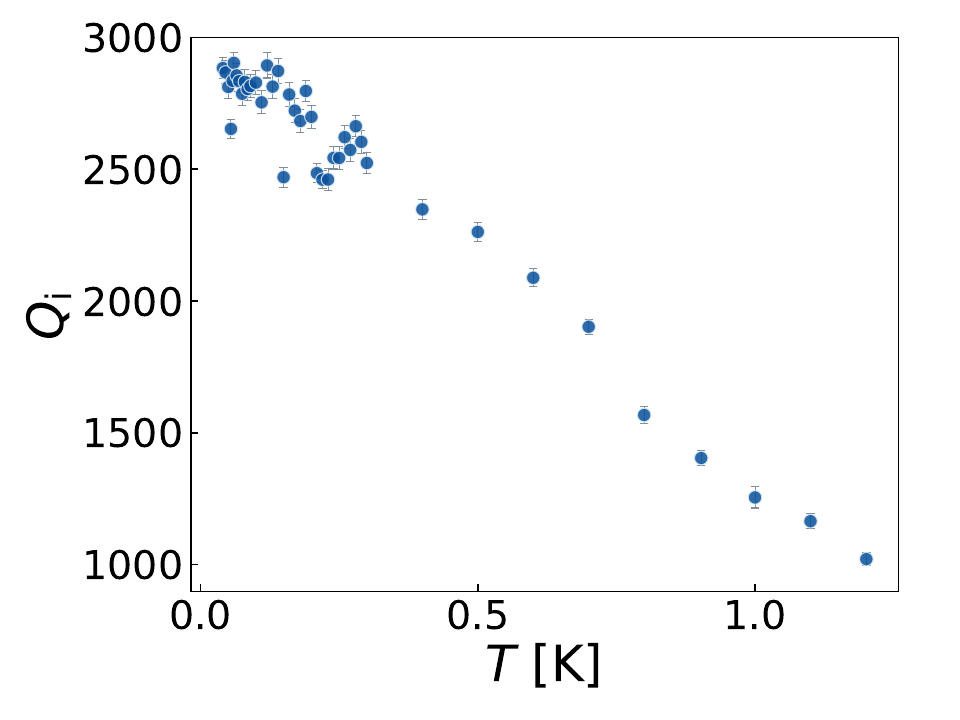}
   \caption{Internal quality factor as a function of temperature of the hybrid device, discussed in the main text.}
   \label{fig:Qi}
\end{figure*}

We use a general fitting function to analyze the complex transmission ($S_{\text{21}}$) of a notch type resonator [68,81]: %\cite{gao2008physics,probst2015efficient} :

\begin{align}
S_{21}(f)= 1- \frac{\left(Q_{tot} /\left|Q_{c}\right|\right) e^{i \phi}}{1+2 i Q_{tot}\left(f / f_{r}-1\right)}
\end{align}

where $f$ denotes the measured frequency, $f_{r}$ the resonance frequency, and $\left|Q_{c}\right|$ the absolute value of the coupling quality factor, and $\phi$ quantifies the impedance mismatch. The total quality factor $Q_{tot}$ is given by $Q_{tot}=1/(Q_i^{-1}+Q_c^{-1})$, where $Q_i$ is the internal quality factor.

We extract the internal quality factor from measured $S_{\text{21}}$ of a hybrid device, discussed in the main text (Fig.4). Figure \ref{fig:Qi} shows the temperature dependence of the internal quality factor $Q_i$. As the temperature increases, $Q_i$ exhibits an approximately linear decrease over the measured temperature range, indicating a corresponding increase in microwave dissipation. In contrast to the resonance frequency shift, which is primarily sensitive to changes in the kinetic inductance and therefore probes the imaginary part of the complex conductivity, $Q_i$ is governed by dissipative processes associated with the real part of the conductivity. Because $Q_i$ reflects the combined contributions of quasiparticle dissipation and other loss mechanisms within the hybrid structure, its temperature dependence provides a less direct probe of the superconducting gap structure than the resonance frequency shift. Accordingly, the main text focuses on the temperature dependent resonance frequency, whose linear variation is consistent with a nodal superconducting gap. Nevertheless, the approximately linear decrease of $Q_i$ with temperature is qualitatively consistent with the presence of thermally excited low-energy quasiparticles expected for a nodal superconducting state.

\vspace{2em}
\section{E\MakeLowercase{ffect of interface inhomogeneity on measured resonance frequency }}  

\begin{figure*}[h!]
   \centering
   \includegraphics[width=0.87\textwidth]{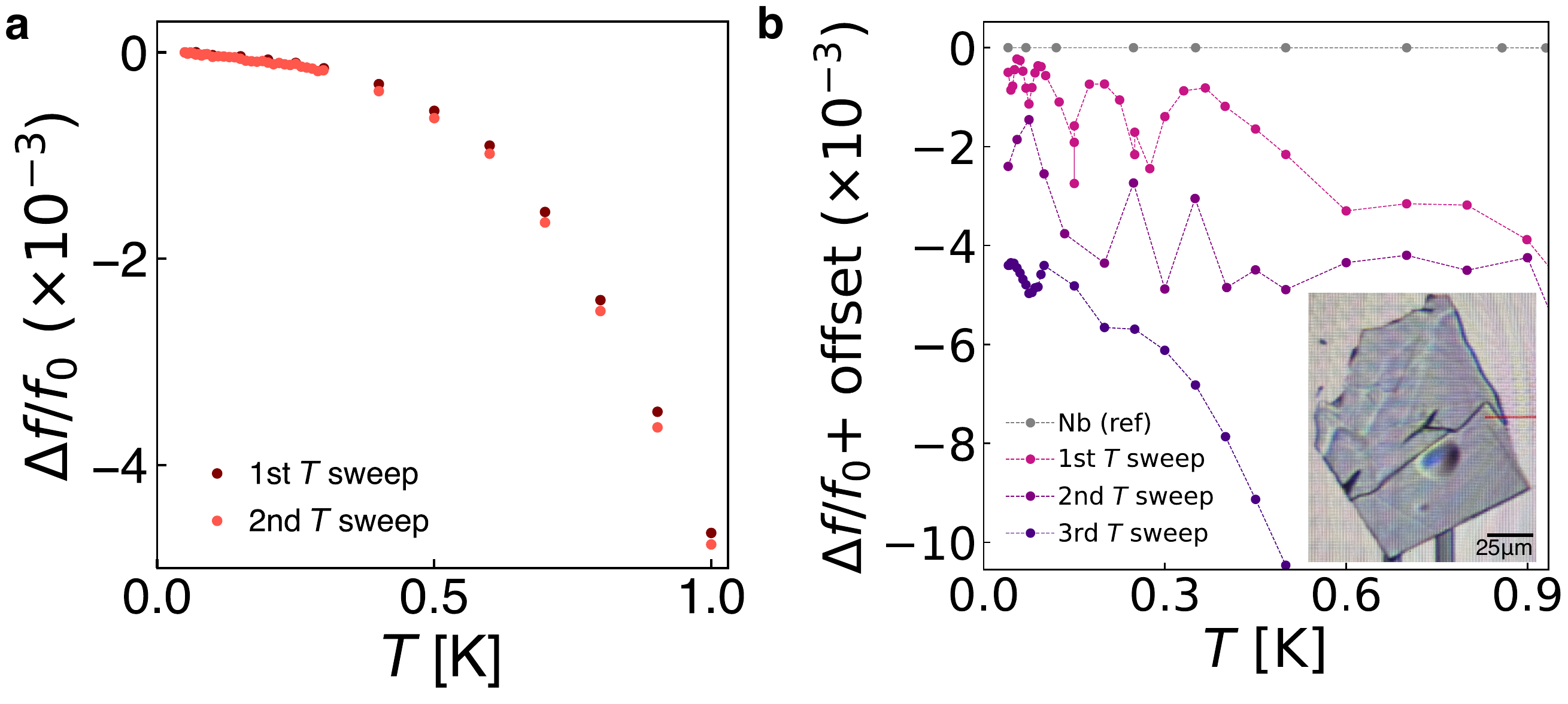}
   \caption{a) Temperature dependent resonance frequency shift for the hybrid device presented in the main text. Repeated temperature sweep shows consistent linear behavior. b) Temperature-dependent resonance frequency of a device exhibiting a non-uniform interface between the flake and the resonator (bubble formation visible in inset.) Repeated temperature sweeps show significant non-reproducibility and fluctuations. A third sweep, performed after higher microwave power was applied, results in a partially linear trend with a additional structure. }
   \label{fig:bubble}
\end{figure*}
We measure the temperature dependence of the resonance frequency in a hybrid device exhibiting a visible interface inhomogeneity (bubble) between a flake and a resonator film. In contrast to the devices presented in the main text (see Fig. \ref{fig:bubble}a), repeated temperature sweeps on this sample show pronounced non-reproducibility and fluctuations in the resonance frequency, as shown in Fig. \ref{fig:bubble}b. We attribute this behavior to the non-uniform interface, which likely leads to unstable coupling, and additional extrinsic contributions. 
Moreover, the imperfect interface may perturb current distribution and effective capacitance. During a subsequent sweep, higher microwave power was applied before cooling for the third sweep, likely modifying the local interface conditions (e.g., via heating or defect rearrangement). While the resulting data exhibits a partially linear temperature dependence, significant additional structure remains, indicating that the extracted response is still influenced by extrinsic effects.
These observations emphasize that, in the presence of imperfect interface conditions, the intrinsic temperature dependence cannot be reliably extracted. 
In this device, the resonator geometry was also modified with decreased capping layer thickness and an increased isolation gap, which lowers the characteristic capacitance and can influence two level system coupling. The reference resonator exhibits a quality factor of approximately 2.5 $\times10^4$. While these design changes may contribute to an overall degradation of resonator performance, they do not account for the pronounced non-reproducibility and fluctuations observed across repeated temperature sweeps. We therefore attribute the dominant source of inconsistent resonance frequency response to the non-uniform interface (bubble formation), which introduces uncontrolled, temperature dependent extrinsic effects.
These observations highlight the critical importance of a clean and uniform interface for obtaining reliable temperature dependence resonance shifts.

\section{Superfluid density extraction}

To characterize the superconducting properties of the flake, we extract the superfluid density from the measured resonance frequency of the device, which exhibits a linear decrease at low temperatures, as shown in the main text. The temperature dependence of the resonance frequency reflects variations in the kinetic inductance of the flake. By adapting the transmission line resonator model (described in section III), and relating the observed frequency shifts to changes in kinetic inductance, we can determine the temperature-dependent superfluid density. % providing insight into the superconducting gap structure. 

\subsection{Expansion near the \(\lambda/4\) mode}
For  modes close to the quarter-wavelength resonance, the propagation constant can be written as 
\begin{equation}
\beta l=\frac{\pi}{2}+\delta, \qquad |\delta|\ll 1,
\end{equation}
with $\delta$ representing a small shift due to the terminating load. Using the approximation $\tan(\pi/2+\delta) \approx -1/\delta$ in the general resonance condition, we find
\begin{equation}
\delta
\approx
-\frac{\omega L_k-\dfrac{1}{\omega C_e}}{Z_0}.
\end{equation}
The corresponding resonance frequency is then
\begin{equation}
\omega
\approx
\omega_{1/4}+\frac{v_p}{l}\,\delta,
\qquad
\omega_{1/4}=\frac{\pi v_p}{2l},
\end{equation}
giving a frequency shift due to kinetic inductance: 
\begin{equation}
\Delta\omega
\approx
-\frac{v_p}{lZ_0}
\left(\omega L_k-\frac{1}{\omega C_e}\right).
\end{equation}
Then, the fractional shift in kinetic inductance leads to:
\begin{equation}
\frac{\Delta\omega(T)}{\omega_0}
\approx
-\frac{v_p}{lZ_0}\Delta L_k(T),
\qquad
\Delta L_k(T)= L_k(T) -L_{k0}
\end{equation}
Converting to $f=\omega/2\pi$, the fractional shift becomes
\begin{equation}
\frac{\Delta f(T)}{f_0}
\approx
-\frac{v_p}{lZ_0}\Delta L_k(T).
\end{equation}

\subsection{Kinetic inductance and superfluid density}

Since $L_{\text{k}}\propto1/n_{s}(T)$, we define 
\begin{equation}
    L_{k}(T) = L_{k0} \frac{n_s(0)}{n_s(T)},
    \qquad
    \Delta L_{k}(T) = L_{k0} \left(  \frac{n_s(0)}{n_s(T)} -1\right)
\end{equation}

so that
\begin{equation}
    \frac{n_s(T)}{n_s(0)} \approx \frac{1}{1+ \frac{l Z_0}{v_p L_{k0}} \frac{-\Delta f(T)}{f_0}}
\end{equation}

\subsection{Low-temperature linear approximation}

We adopt the low-temperature expression for the normalized superfluid density derived for $d$-wave cuprate superconductors [75, 76]:%\cite{lee1997unusual, sutherland2003thermal}:
\begin{equation}
\frac{n_s(T)}{n_s(0)}
=
1
-
\left[
\frac{2\ln 2\, \mu_0 e^2}{\pi}
\frac{\lambda_0^2}{\hbar^2}
\frac{n}{d}
\frac{v_F}{v_\Delta}
\right]
k_B T,
\qquad (T \ll \Delta_0/k_B),
\end{equation}

where $\mu_0$ is the vacuum permeability, $e$ the elementary charge, $\lambda_0$ the zero-temperature London penetration depth, $\hbar$ the reduced Planck constant, and $k_B$ the Boltzmann constant. The parameter $n$ denotes the number of conducting layers per unit cell, $d$ is the interlayer spacing (such that $n/d$ gives the number of conducting layers per unit length). 
Here, $v_F$ is the Fermi velocity normal to the Fermi surface at the node, $v_{\Delta}$ is the gap velocity along the Fermi surface ($v_{\Delta} \equiv \left. \partial \Delta(k)/\partial k_{\parallel} \right|_{\mathrm{node}}$), and $k_{\parallel}$ is the momentum tangent to the Fermi surface. 

While our system may not host a pure $d$-wave gap structure, this expression is used as an approximate analogy to provide qualitative insight into the low-energy behavior of the superfluid density.

\begin{table}[h]
\centering
\begin{tabular}{cccccc}
\hline\hline
Reference & $\lambda_0$ & $v_F$ & $v_{\Delta}$ & $k_F$ \\
 & (nm) & (eV$\cdot$\AA) & (meV$\cdot$\AA) & (\AA$^{-1}$) \\
\hline

Critical magnetic fields measurement [77] %\cite{ni2021anisotropic}
&  460
& - 
& - 
& - \\
Tunneling diode oscillator [78]& 58 (calculation)& - %\cite{grant2025superconducting}
& - 
& - \\
Tunneling diode oscillator [48]%\cite{duan2021nodeless}
&  387
&  -
&  -
&  - \\
µ-SR [50]%\cite{gupta2022microscopic}
&  258($\lambda_{ab}$), 949($\lambda_{c}$) (s+s-wave model)
& -
& -  
& - \\

&  249($\lambda_{ab}$), 887($\lambda_{c}$) (d-wave model)
& 
&   
&  \\
SQUID susceptometry [54]% \cite{kaczmarek2025direct}
&  195-390
&  - 
&  -
&  -\\ 

ARPES [79]%\cite{zhong2023testing}
&  -
& 3.0
& - 
& 0.5 $\pm~0.01$\\ 

ARPES [56]%\cite{mine2025observation}
&  -
& 2.6 
& 2.3 
& 0.5 $\pm~0.02$ \\ 
\hline\hline
\end{tabular}
\caption{Parameters used to estimate the low-temperature superfluid-density slope. Here, we adopt the Fermi velocity $v_F$ based on the slope of the band dispersion of the hexagonal pocket ($\beta$ Fermi surface) along the $\Gamma \text{K}$ direction [56,79]. %\cite{zhong2023testing,mine2025observation} 
The gap velocity $v_{\Delta}$ is obtained from the slope of the angle-dependent superconducting gap magnitude on the $\beta$ Fermi surface [56].} %\cite{mine2025observation}.}
\label{tab:penetration depth and fermi velocity}
\end{table}

To obtain a quantitative estimate, we adopt the $v_f$, $v_\Delta$, and $\lambda_0$ from previous experiments on this material, listed in table \ref{tab:penetration depth and fermi velocity}. Using these values and taking the layer density $n/d=0.1 ~\mathrm{\AA}^{-1}$, the normalized superfluid density is expected to follow
\begin{equation}
\frac{n_s(T)}{n_s(0)} \approx 1 - (0.067\sim 5.534)\,T,
\end{equation}
where the slope range reflects the variation in the effective parameters from literature. In our device (Fig.~4(d) in the main text), the experimentally extracted superfluid density at low temperatures exhibits
\begin{equation}
\frac{n_s(T)}{n_s(0)} \approx 1 - 0.04\,T.   %0-0.4K
\end{equation}
While this estimate is based on an analogy to $d$-wave cuprates and does not provide a precise determination of the superconducting gap structure, it allows a qualitative comparison of the measured low temperature linear behavior with an effective scale of low-energy excitations.

{}